\documentclass[12pt]{article}
\usepackage{graphicx}
\begin{document}

\def \d {{\rm d}}

\title{The extensions of gravitational soliton solutions with real poles}

\author{J. B. Griffiths\thanks{E--mail: {\tt J.B.Griffiths@Lboro.ac.uk}} \ 
and S. Miccich\`e\thanks{E--mail: {\tt S.Micciche@Lboro.ac.uk}} \\ \\
Department of Mathematical Sciences, Loughborough University, \\
Loughborough, Leics., LE11 3TU, U.K. \\ }

\maketitle

\baselineskip=14pt

\begin{abstract} 
\noindent We analyse vacuum gravitational ``soliton'' solutions with real
poles in the cosmological context. It is well known that these solutions 
contain singularities on certain null hypersurfaces. Using a Kasner seed
solution, we demonstrate that these may contain thin sheets of null matter or
may be simple coordinate singularities, and we describe a number of possible
extensions through them. 
\end{abstract}

\section {Introduction}

The inverse scattering (BZ) technique of Belinskii and Zakharov
\cite{BelZak78} is now well known. It is essentially a solution-generating
procedure for producing exact vacuum solutions of Einstein's equations for
space-times admitting two isometries. Starting from some initial ``seed''
solution, the technique is based on the construction of a ``dressing''
matrix which is a meromorphic function of a complex spectral parameter
$\lambda$.

For the case in which a vacuum space-time admits two hypersurface-orthogonal
spacelike Killing vectors, the ``gravitational solitons'' corresponding to
particular poles of the dressing matrix generally describe perturbations of
the gravitational field which propagate like finite gravitational waves on
some background. Here we consider gravitational soliton solutions in such
space-times which correspond to real poles in the scattering matrix. As
originally pointed out by Belinskii and Zakharov \cite{BelZak78}, these
solutions exist in regions which are bounded by null hypersurfaces on which
singularities occur. It is the purpose of this paper to reconsider the
character of these singularities and the possible extensions through them.

Carr and Verdaguer \cite{CarVer83} have considered soliton solutions in a
Kasner background and have interpreted the solutions with real poles as
inhomogeneous cosmologies with shock waves in which the solitons propagate
away revealing the Kasner background. However, as shown by Gleiser
\cite{Gle84} and Curir, Francaviglia and Verdaguer \cite{CuFrVe92}, these
solutions must contain thin sheets of null matter separating the various
regions. Gleiser \cite{Gle84} has also described alternative matter-free
extensions, while Curir, Francaviglia and Verdaguer \cite{CuFrVe92} have
considered a real pole of arbitrary degeneracy for the diagonal ``soliton''
solution of Carmeli and Charach \cite{CarCha80} which was shown to correspond
to a real (degenerate) pole by Feinstein and Charach \cite{FeiCha86}.

Others \cite{BelFra82}--\cite{BraCur93} have considered soliton solutions
with real poles in which the seed metric is a nondiagonal vacuum Bianchi~II
space-time. In this case, it is similarly possible to remove the coordinate
singularity on the null hypersurface, although there may again be an
impulsive gravitational wave and a $\delta$-function in the Ricci tensor on
the shock front.

In this paper, we reconsider the physical interpretation of some soliton
solutions with real poles. As summarised above, these can represent
gravitational shock waves in some cosmological background. However, the
extension to the background is not unique even if the possibility of thin
sheets of null matter is excluded.  In particular, we consider the one-soliton
solution with a vacuum Bianchi~I seed. By extending this through the shock
front in various ways, we construct a number of different global solutions.

\section {Real pole solitons with a Kasner seed}

According to the BZ technique \cite{BelZak78}, we consider a vacuum
space-time with two hyper\-surface-orthogonal Killing vectors. In the case in
which the isometries are spacelike, the metric can be written in the form 
 \begin{equation}
 \d s^2=2e^{-M}\d u\,\d v -g_{ij}\d x^i\d x^j 
 \end{equation}
 where the 2-metric {\bf g} $(=g_{ij})$ is a function of the two null
coordinates $u$ and $v$, and has the determinant $|{\bf g}|=\alpha^2$.
Einstein's vacuum field equations require that $\alpha$ satisfies the
2-dimensional wave equation and so can be written in the form \
$\alpha=f(u)+g(v)$, \ where $f(u)$ and $g(v)$ are arbitrary functions. It is
also convenient to introduce another function $\beta(u,v)$ which is
harmonically conjugate to $\alpha$ and given by $\beta=f(u)-g(v)$.  In this
and following sections, we will assume that the two null coordinates $u$ and
$v$ are future-pointing.

Here, we start with an initial Kasner seed solution (denoted by a suffix zero)
which can be written in the form (1) with 
 \begin{equation}
 {\bf g}_0=\alpha\pmatrix{ \alpha^p &0 \cr 0 &\alpha^{-p} \cr}, \qquad
e^{-M_0}={\alpha_u\alpha_v\over\alpha^{(1-p^2)/2}} 
 \end{equation}
 where $p$ is an arbitrary parameter. This reduces to a form of the
Minkowski metric when $p=\pm1$. According to the BZ technique, we work with
the matrix 
 $$ \Psi_0(u,v,\lambda)=\pmatrix{
(\alpha^2+2\beta\lambda+\lambda^2)^{(1+p)/2} &0 \cr 
\noalign{\smallskip}
0 &(\alpha^2+2\beta\lambda+\lambda^2)^{(1-p)/2} \cr} $$ 
 which satisfies the appropriate equation and the condition
$\Psi_0(u,v,0)={\bf g}_0$.

We also restrict attention here to the case in which there is a single real
pole given by $\lambda=\mu$, where
 $$ \mu=\omega-\beta+\sqrt{(\omega-\beta)^2-\alpha^2} $$
 and $\omega$ is an arbitrary real constant. (For a single real pole, we note
that the alternative expression \
$\mu=\omega-\beta-\sqrt{(\omega-\beta)^2-\alpha^2}$ \ simply corresponds to a
rotation of coordinates $x\to y$, $y\to-x$.) Clearly the single soliton
solution is only admissible in regions of space-time for which
$(\omega-\beta)^2\ge\alpha^2$. (If there are more than one such regions,
these will normally be disjoint.) These regions will be bounded by an initial
cosmological curvature singularity which occurs when $\alpha=0$. We will
therefore take $\alpha>0$. They will also be bounded by null hypersurfaces on
which $(\omega-\beta)^2=\alpha^2$. According to common usage, we will refer to
these as ``shock fronts''.

Since $\alpha^2+2\beta\lambda+\lambda^2=2\omega\lambda$, we obtain that on the
pole trajectory $\lambda=\mu$  
 $$ \Psi_0^{-1}(u,v,\mu) =\pmatrix{ (2\omega\mu)^{-(1+p)/2} &0 \cr
\noalign{\smallskip}
0 &(2\omega\mu)^{-(1-p)/2} \cr }. $$ 
 The procedure is now well known and, for the nondiagonal case in the region
$\omega-\beta\ge\alpha$, the new solution can be expressed as 
 $$ {\bf g}=\left({\mu\over\alpha}{\bf I} 
-{(\mu^2-\alpha^2)\over\alpha\mu}{\bf P}\right){\bf g}_0 $$ 
 where, after introducing a new function $s(\alpha,\beta)$ such that
$e^s=\mu/\alpha$ and the constant $c=p\log(2\omega)$, the matrix {\bf P} is
given by
 $$ {\bf P}={1\over2\cosh(ps+c)} \pmatrix{ e^{-ps-c} &\alpha^p \cr
\noalign{\smallskip}
\alpha^{-p} &e^{ps+c} \cr }. $$ 
 It may be noted that a change in the values of the free soliton parameters
corresponds to a rotation of coordinates. For the single soliton solution,
this is not physically significant. However, when attached to other regions,
changes in these parameters can represent soliton waves with different
polarization. With the above expressions, the new solution is given by 
 \begin{equation}
 {\bf g}= {\alpha\over\cosh(ps+c)} 
\pmatrix{ \alpha^p\cosh[(1+p)s+c] &-\sinh s \cr
\noalign{\smallskip}
-\sinh s &\alpha^{-p}\cosh[(1-p)s-c] \cr}. 
 \end{equation}

\section {Extending the solution to a Kasner background}

As noted, the above solution using (3) is only defined in the region of
space-time in which $\omega-\beta\ge\alpha>0$. This region is bounded by the
null hypersurface $\beta+\alpha=\omega$ on which $\mu=\alpha$. It can be seen
that, on this boundary, $s=0$ and the 2-metric is the same as that of the
seed. Thus, it would seem to be possible to join this metric continuously
with the seed Kasner solution in the region $\alpha\ge\omega-\beta$, thus
forming a gravitational shock front as described by Carr and Verdaguer
\cite{CarVer83}. It is also possible to include a one-soliton solution in the
region in which $\beta-\omega\ge\alpha>0$. The standard global interpretation
of this solution is then as two disjointed one-soliton regions moving apart
leaving an exact Kasner background. This forms a composite space-time as
illustrated in figure~1 in which there is an initial curvature singularity when
$\alpha=0$.

However, although the 2-metric is continuous in this construction, it may have
discontinuous first derivatives which could introduce an impulsive
gravitational wave component on the shock front.

More seriously, we note that, using (3), the new expression for $M$ in the
region $\omega-\beta\ge\alpha>0$ is given by
 $$ e^{-M}= {C\,\sqrt\alpha\,\cosh(ps+c)\over 
\sqrt{\omega-\alpha-\beta}\sqrt{\omega+\alpha-\beta}} \,e^{-M_0} $$ 
 where $C$ is an arbitrary constant. This clearly introduces a singularity on
the shock front $\beta+\alpha=\omega$. However this can be seen to be a
coordinate singularity which can be removed by a particular choice of the
functions $f(u)$. For example, using (2), the singularity can be removed by
the choice $f={1\over2}(\omega-u^2)$.

\begin{figure}
\begin{center} \includegraphics[scale=0.5, trim=5 5 5 5]{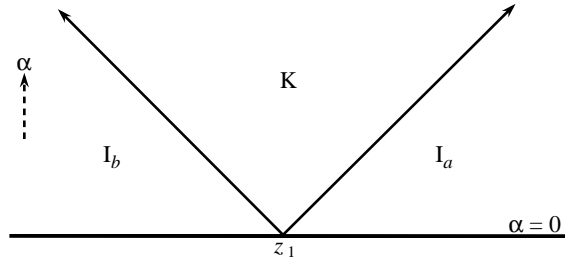}
\caption{ This represents a one-soliton solution, using $\alpha$ and $\beta$
as coordinates, in which the solitons propagate away to reveal a Kasner
background. The space-time is a combination of two one-soliton regions
(denoted by I$_a$ and I$_b$) each with a real pole $\omega_1=-z_1$, together
with a Kasner region (denoted by K). } 
\end{center}
\end{figure}

It may be recalled that the BZ method is based on the assumption that
$\alpha=f+g$ is the same for the soliton solution as for the seed metric that
was used to generate it. Thus, with the choice $f={1\over2}(\omega-u^2)$, the
seed that is used in this case is a particular form of the Kasner metric which
contains a coordinate singularity on the null hypersurface~$u=0$. \ i.e. in
order to generate a non-singular one-soliton solution directly, it is
necessary to start with a seed solution in a form containing a coordinate
singularity.

In the above construction, the space-time has been extended through the null
hypersurface $u=0$, on which $\mu=\alpha$, to a ``background'' region which is
part of a Kasner space-time with the same parameter $p$ as the seed solution.
However, to avoid a coordinate singularity in this region, it is necessary to
put $\alpha={1\over2}(u+v)$, which is equivalent to the choice \
$f={1\over2}(u+\omega)$ and $g={1\over2}(v-\omega)$. \ In order for this to be
continuous with a nonsingular soliton region, it is necessary for the soliton
region to have \ $f={1\over2}(\omega-u^2)$ and $g={1\over2}(v-\omega)$ \
(with the region I$_b$ the same but with $u$ and $v$ interchanged). From this,
it follows that $\alpha$ in the soliton regions (and the associated seed) is
different to that in any extended ``background'' region, and that the
background only has the same form as the seed solution after a coordinate
transformation.

In the soliton region I$_a$ we now have $\alpha={1\over2}(v-u^2)$. This has
been made continuous with a Kasner background in which $\alpha={1\over2}(u+v)$
across $u=0$. However, it may be recalled that discontinuities in the
derivatives of $\alpha$ across a hypersurface induce nonzero components in
the Ricci tensor, and hence in the energy-momentum tensor. In this case, these
are given by  
 $$ T_{uu}=-{e^M\over8\pi\alpha}\,[\alpha_u]\delta(u) \qquad {\rm or} \qquad
T_{vv}=-{e^M\over8\pi\alpha}\,[\alpha_v]\delta(v). $$ 
 It can thus be seen that the discontinuity in the derivative of $\alpha$
across $u=0$ in the above expressions gives rise to an impulsive component in
the energy-momentum tensor corresponding to a thin sheet of null matter
located on this hypersurface. This has been described explicitly elsewhere
\cite{Gle84},~\cite{CuFrVe92}. Moreover, with this time orientation, the
matter has {\it negative} energy density~\cite{BraCur93}.

We now have a one-soliton solution (3) with a single real pole
$\omega_1=-z_1$ in the region represented as I$_a$ in figure~1. This has been
extended to a Kasner region (represented by K in figure~1) across the shock
front $u=-z_1$, and another one-soliton region (represented by I$_b$) in which
$\omega-\beta<-\alpha$ has been added together with a second shock front
$v=z_1$. The metric function $\alpha=f(u)+g(v)$ is determined by the following
expressions in each region:

\bigskip
\begin{tabular}{lccll}

K: &$u>-z_1$, & $v>z_1$ &\qquad
$f={1\over2}u$ &\quad $g={1\over2}v$\vspace{10pt}\\

I$_a$: & $u\le-z_1$, & $v>z_1$ &\qquad
$f={1\over2}\big[-z_1-(u+z_1)^2\big]$ &\quad
$g={1\over2}v$\vspace{10pt}\\

I$_b$: & $u>-z_1$,& $v\le z_1$ &\qquad
$f={1\over2}u$ &\quad
$g={1\over2}\big[z_1-(z_1-v)^2\big]$\\

\end{tabular}

\bigskip\noindent
Clearly there are discontinuities in the derivatives of $\alpha$ across the
junctions $u=-z_1$ and $v=z_1$. Thus, as described above, this configuration
must contain thin sheets of null matter on these hypersurfaces.

\begin{figure}
\begin{center} \includegraphics[scale=0.5, trim=5 5 5 5]{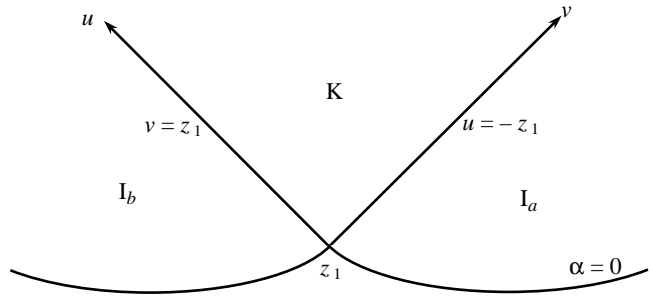}
\caption{ This represents the one-soliton solution in which the solitons
propagate away to reveal a Kasner background in the $u,v$ coordinates. An
initial ``cosmological'' curvature singularity occurs when $\alpha=0$. } 
\end{center}
\end{figure}

It may also be noted that the curvature singularity at $\alpha=0$ is located,
in these coordinates, on $v-z_1-(u+z_1)^2=0$ for $v>z_1$ and on
$u+z_1-(z_1-v)^2=0$ for $v\le z_1$. Thus, although this singularity is
spacelike, it approaches a null limit at the junction point $v=-u=z_1$. This
is illustrated in figure~2.

It is appropriate to work with a null tetrad such that $\ell_i=e^{-M/2}u_{,i}$
and $n_i=e^{-M/2}v_{,i}$. Since $M$ is continuous across the shock fronts in
these coordinates, these null vectors are well behaved throughout the
space-time. Using these, it can then be shown that (provided $p\ne0,\pm1$) the
Weyl tensor has non-zero components $\Psi_0$, $\Psi_2$ and $\Psi_4$ in both
the soliton and Kasner regions. Moreover, these components are bounded near
the shock fronts, although they are not necessarily continuous across the
shock which may also have step changes in the polarization into the soliton
regions (in the nondiagonal case). In addition, the shock fronts themselves
may also contain impulsive components.

\section {Soliton solutions with distinct real poles}

In this section we re-consider the $n$-soliton solutions with distinct real
poles which have been described briefly elsewhere \cite{CarVer83} and
interpreted as solitons moving apart leaving an exact Kasner background. To be
specific, we concentrate on the case of a soliton solution with two distinct
real poles $\omega_1=-z_1$ and $\omega_2=-z_2$, where $z_2>z_1$. This can
again be interpreted as a composite space-time having one- and two-soliton
regions and a Kasner background as illustrated in figure~3 using
$\alpha,\beta$ coordinates. In this case, the shock fronts can be taken to be
$u=-z_1$, $u=-z_2$, $v=z_1$ and $v=z_2$.

As described in the previous section, singularities can occur in the metric
coefficient $e^{-M}$. For the two-soliton solution this is given by 
 $$ e^{-M} ={C \>\alpha^{((2-p)^2-1)/2}
(\mu_1\mu_2)^{2+p} \> \alpha_u\alpha_v \over
(\alpha^2-\mu_1^2)(\alpha^2-\mu_1\mu_2)^2(\alpha^2-\mu_2^2)}. $$ 
 An explicit transformation which removes each singularity individually has
been given by D\'{\i}az and Gleiser \cite{DiaGle88} (for the general case of
$n$ distinct real poles). However, this can only be applied to one singularity
at a time as the resulting metric becomes discontinuous on the next singular
wavefront. Below, we introduce a different gauge in which the metric is
continuous throughout the space-time.

\begin{figure}
\begin{center} \includegraphics[scale=0.65, trim=5 5 5 5]{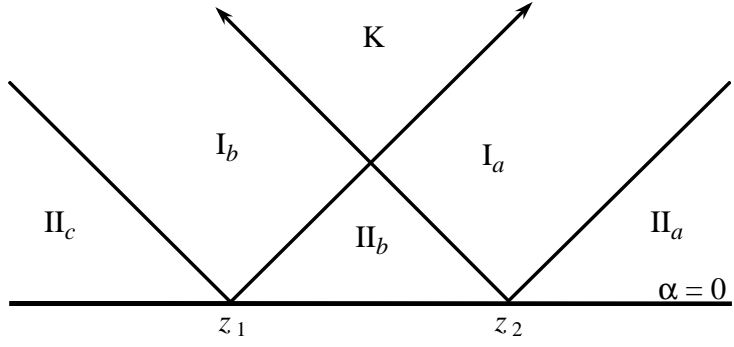}
\caption{ This represents soliton solutions with two real poles with $\alpha$
and $\beta$ as coordinates. The solitons propagate away to reveal a Kasner
background. The space-time is a combination of one- and two-soliton regions
(denoted by I$_a$, I$_b$, II$_a$, II$_b$, II$_c$) with real poles
$\omega_1=-z_1$ and $\omega_2=-z_2$, together with a Kasner region (denoted by
K). } 
\end{center}
\end{figure}

The two-soliton solution is appropriate in the regions II$_a$, II$_b$ and
II$_c$ indicated in figure~2. It is then possible to extend the solution into
regions described by the one-soliton solution (3) and then further to the
Kasner solution (2) as illustrated. We can adopt two future-pointing null
coordinates $u$ and $v$ that are defined globally. It is then possible to
introduce a gauge such that $f(u)$ and $g(v)$, and hence $\alpha(u,v)$ and
$\beta(u,v)$ are continuous everywhere. One such gauge is as follows: 

\bigskip

\begin{tabular}{lccl}

K:& $u>-z_1$,& $v>z_2$, &\qquad
$\left\{\begin{array}{l}
f={1\over2}u,\\
\noalign{\smallskip}
g={1\over2}v
\end{array}\right.$\vspace{10pt}\\

I$_a$:& $-z_1\ge u\ge-z_2$,& $v>z_2$,&\qquad
$\left\{\begin{array}{l}
f={1\over2}\big[-z_1-(u+z_1)^2-k(u+z_1)^3\big],\\
\noalign{\smallskip}
g={1\over2}v
\end{array}\right.$
\vspace{10pt}\\

I$_b$:& $u>-z_1$,& $z_2\ge v>z_1$, &\qquad
$\left\{\begin{array}{l}
f={1\over2}u,\\
\noalign{\smallskip}
g={1\over2}\big[z_2-(z_2-v)^2+k(z_2-v)^3\big]
\end{array}\right.$
\vspace{10pt}\\

II$_a$:& $-z_2\ge u$,& $v>z_2$, &\qquad
$\left\{\begin{array}{l}
f={1\over2}\big[-z_2-(u+z_2)^2\big],\\
\noalign{\smallskip}
g={1\over2}v
\end{array}\right.$
\vspace{10pt}\\

II$_b$:& $-z_1\ge u\ge-z_2$,& $z_2\ge v>z_1$, &\qquad
$\left\{\begin{array}{l}
f={1\over2}\big[-z_1-(u+z_1)^2-k(u+z_1)^3\big],\\
\noalign{\smallskip}
g={1\over2}\big[z_2-(z_2-v)^2+k(z_2-v)^3\big]
\end{array}\right.$
\vspace{10pt}\\

II$_c$:& $u>-z_1$,& $z_1\ge v$, &\qquad
$\left\{\begin{array}{l}
f={1\over2}u,\\
\noalign{\smallskip}
g={1\over2}\big[z_1-(z_1-v)^2\big]
\end{array}\right.$
\vspace{10pt}\\

\end{tabular}

\bigskip\noindent
where $k=(z_2-z_1)^{-1}-(z_2-z_1)^{-2}$.

\goodbreak
With these expressions, it can be seen that the metric is $C^0$ everywhere
and that all coordinate singularities have been removed. However, it can also
be seen that there are discontinuities in the derivatives of $\alpha$ on the
null hypersurfaces $u=-z_1$, $u=-z_2$, $v=z_1$ and $v=z_2$ which therefore
must contain thin sheets of null matter. The structure in $u,v$ coordinates is
illustrated in figure~4.

\begin{figure}
\begin{center} \includegraphics[scale=0.65, trim=5 5 5 5]{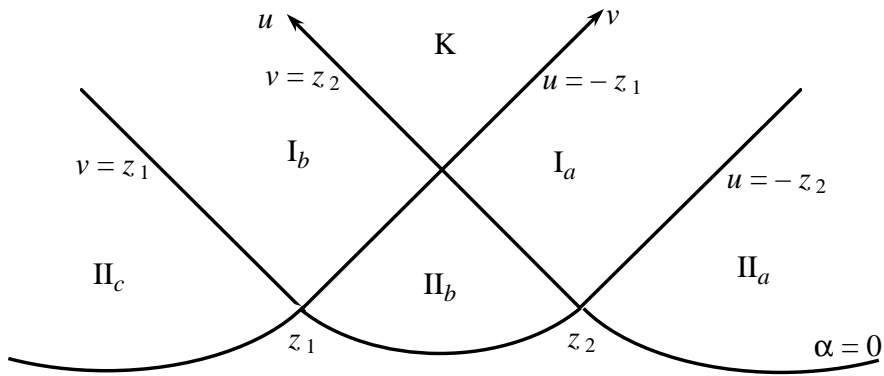}
\caption{ This represents soliton solutions with two real poles in the $u,v$
coordinates. An initial ``cosmological'' curvature singularity occurs when
$\alpha=0$. } 
\end{center}
\end{figure}

It may be noted that the space-times in the soliton regions may be diagonal or
non-diagonal. However, some soliton parameters must be continued across the
shock fronts. To be explicit, the one-soliton regions must contain solitons
whose parameters are continued from the two-soliton regions II$_a$ or II$_c$.
The region II$_b$, however, must contain a continuation of the two solitons
which extend back from regions I$_a$ and I$_b$ and therefore its parameters
are predetermined from those of II$_a$ and II$_c$.

Clearly, this approach can be generalised to include an arbitrary number of
distinct real poles. For solutions with more than one real pole, these
solutions necessarily contain thin sheets of null matter with negative energy
density on the boundaries between the various regions.

\section {Possible extensions without sheets of null matter}

We now consider the possible extensions of the one-soliton solution which do
not involve thin sheets of null matter. In this case, no generality is lost
in making a coordinate shift to put $\omega=0$ so that the shock front of the
soliton region is then given by $u=0$. We also continue to use two
future-pointing null coordinates $u$ and $v$.

We may initially consider whether or not it is possible to construct an exact
one-soliton solution which has the same global structure as that indicated in
Figure~1, but without the presence of thin sheets of null matter. For this, the
regions $I_a$ and $I_b$ would be essentially the same, but the extension would
not be to a Kasner background. If, in the one-soliton region $I_a$ we put
$\alpha={1\over2}(v^2-u^2)$, it may be possible in the extended region to
choose a gauge such that $\alpha={1\over2}(u^2+v^2)$. However, it is shown in
the appendix that such an extension is not possible. We therefore look for
alternative extensions to the region~I$_a$. Since the shock front is null, any
extension will be non-unique. In fact, a number of possibilities are readily
available.

It may first be noted that the one-soliton region is algebraically general.
There will therefore be a gravitational wave component propagating towards and
through the shock front. In the soliton region, there will also be a
gravitational wave component propagating in the opposite direction parallel to
the shock front. However, this component need not occur in the extended region
and, in this case, the extension will be to a plane wave region. Other
possibilities include that in which the extension is to another distinct
soliton region, or to that containing an arbitrary gravitational wave
component propagating parallel to the shock front (although only the linear
case will be constructed below).

\subsection {A plane wave extension}
The simplest extension across the shock front $u=0$ is to a plane wave region
in which $f=0$ and {\bf g} and $M$ are functions of $v$ only. The general
structure is illustrated in figure~5.

\begin{figure}
\begin{center} \includegraphics[scale=0.5, trim=5 5 5 5]{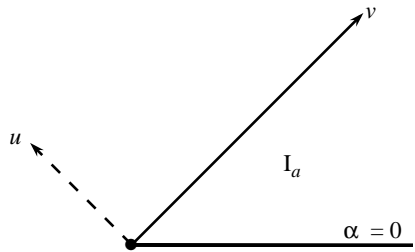}
\caption{ A one-soliton region can be extended to a plane wave region.
Gravitational wave components originating in the singularity propagate into
the plane wave region but, in that region, no wave components propagate along
the lines $u=$~const. A non-scalar curvature singularity occurs on $v=0$. } 
\end{center}
\end{figure}

For the moment, let us continue to adopt a gauge such that $g(v)={1\over2}v$.
In the extension to a plane wave region, we may now set $f(u)=0$ and
$\alpha=-\beta={1\over2}v$ and the metric is given by 
 $$ e^{-M}=c_1v^{(p^2-1)/2}, \qquad 
{\bf g}={v\over2} \pmatrix{ ({v\over2})^p  &0 \cr
\noalign{\smallskip}
0 &({v\over2})^{-p} \cr }. $$ 
 It is then appropriate to make the coordinate transformation 
 $$ v=\tilde v^{2/(p^2+1)} $$ 
 together with a rescaling to remove an unwanted constant so that the line
element becomes 
 $$ \d s^2=2\d u\,\d\tilde v -\tilde v^{2/(p^2+1)} \left(
\tilde v^{2p/(p^2+1)}\d x^2 +\tilde v^{-2p/(p^2+1)}\d y^2 \right). $$ 
 We can then make the further coordinate transformation 
\begin{eqnarray}
X &=& \tilde v^{1+p\over1+p^2}\, x \nonumber\\
Y &=& \tilde v^{1-p\over1+p^2}\, y \nonumber\\
r &=& u +{\textstyle{1\over2}\left({1+p\over1+p^2}\right)}
 \tilde v^{1+2p-p^2\over1+p^2}\, x^2
+{\textstyle{1\over2}\left({1-p\over1+p^2}\right)}
 \tilde v^{1-2p-p^2\over1+p^2}\, y^2 \nonumber
\end{eqnarray}
 to put the line element in the form 
 $$ \d s^2 = 2\d\tilde v\,\d r
-{\textstyle {p(1-p^2)\over(1+p^2)^2}}\, \tilde v^{-2} (X^2-Y^2)\d\tilde v^2
-\d X^2-\d Y^2 $$ 
 which is the familiar form of a plane wave with amplitude profile
$h(\tilde v)={p(1-p^2)\over(1+p^2)^2}\,\tilde v^{-2}$. Thus, except for the
cases in which $p=0,\pm1$ in which the extension is to a flat region, the plane
gravitational wave amplitude in the extended region is clearly unbounded when
$\tilde v=0$ (or $v=0$). This null hypersurface may then reasonably be
considered to form a boundary of the space-time extended from the one-soliton
solution as described in figure~5.

It may be observed that this situation is qualitatively identical to the time
reverse of a colliding plane wave space-time in which a future curvature
singularity is formed following the interaction of initially plane waves. As
in that case, the non-scalar curvature singularity that occurs in the plane
wave region on $v=0$ can be interpreted as a ``fold singularity'' as described
by Matzner and Tipler \cite{MatTip84} (see also \cite{Griff91}).

\subsection {A soliton extension}
An alternative extension can be achieved by matching possibly different
one-soliton solutions on either side of the shock front. On either side we
can adopt the same gauge with \ $f=-{1\over2}\,u^2$ \ and \
$g={1\over2}\,v^2$, \ so that 
 \begin{equation}
 \alpha={\textstyle{1\over2}}(v^2-u^2) \qquad {\rm and} \qquad
\beta=-{\textstyle{1\over2}}(u^2+v^2). 
 \end{equation}
 This clearly satisfies the required inequality for the one-soliton solution.
Taking $u$ to be time-oriented, we now have two regions. An initial region
with $u\le0$ and a second region with $u\ge0$.

It is also appropriate here to introduce alternative coordinates $t$ and
$\rho$ where 
 $$ t={\textstyle{1\over\sqrt2}}(u+v) \qquad {\rm and} \qquad 
\rho={\textstyle{1\over\sqrt2}}(v-u),$$
 so that $\alpha=t\rho$ and $\mu=t^2$.

The curvature singularity at $\alpha=0$ now occurs both when $t=0$ and when
$\rho=0$ so that the space-time is defined only in regions for which $v>|u|$.
However, this includes two regions in which the conditions for a one-soliton
solution are satisfied. These occur on either side of the shock front as
illustrated in figure~6. Of course, there is no reason for solitons in these
two regions to have the same parameters, and it is possible to construct a
compound space-time composed of two different one-soliton solutions joined
across the shock front. For example, we can choose the soliton in one region
to be diagonal, and that in the other non-diagonal. In addition, since $f$ and
$g$ are taken to have the same form in both regions, there is no discontinuity
in the derivatives of $\alpha$, and so there will be no sheets of null matter
across the shock front. This matter-free extension of the one-soliton solution
was initially pointed out by Gleiser~\cite{Gle84} at least for the case when
$p=-1$.

As a particular example, let us consider the case of the one-soliton solutions
that are generated from the plane symmetric (type~D) Kasner solution for which
$p=0$. The non-diagonal case is given by 
 \begin{equation}
 {\bf g}=t\rho \pmatrix{ \cosh s &-\sinh s \cr
\noalign{\smallskip}
-\sinh s &\cosh s \cr } \qquad {\rm where} \qquad e^s=t/\rho. 
 \end{equation}
 This solution can be adopted in the region $t\ge\rho>0$. Choosing
$C=-\omega$, the remaining part of the metric is given by 
 \begin{equation}
 2e^{-M}\d u\,\d v=\d t^2-\d\rho^2. 
 \end{equation}

\begin{figure}
\begin{center} \includegraphics[scale=0.5, trim=5 5 5 5]{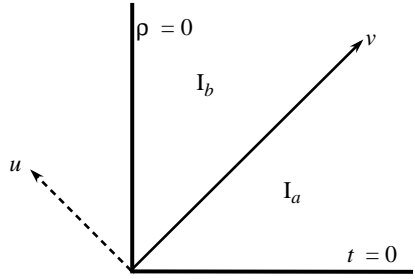}
\caption{ This space-time is composed of two (possibly different) one-soliton
solutions in the two regions denoted by I$_a$, I$_b$. It is defined only for
$v^2-u^2>0$. Generally, there are curvature singularities on the 
hypersurfaces $t=0$ and $\rho=0$. } 
\end{center}
\end{figure}

Let us now attach this to a background region $u<0$ and $v>0$ (i.e.
$\rho>t\ge0$) which is composed of the diagonal one-soliton solution with a
$p=0$ Kasner seed, given by 
 $$ {\bf g}=t\rho \pmatrix{ t/\rho &0 \cr
\noalign{\smallskip}
0 &\rho/t \cr }, \qquad e^{-M}=1. $$ 
 This can be expressed in the form 
 \begin{equation}
 \d s^2=\d t^2-\d\rho^2-t^2\d z^2-\rho^2\d\phi^2 
 \end{equation}
 which after the coordinate transformation 
 $$ T=t\cosh z, \qquad X=\rho\cos\phi, \qquad Y=\rho\sin\phi, 
\qquad Z=t\sinh z $$ 
 is clearly that part $T\ge|Z|$ of the Minkowski space-time 
 $$ \d s^2=\d T^2-\d X^2-\d Y^2-\d Z^2 $$ 
 in cartesian coordinates. In this case, it can be seen that the junction
between the two regions, which is the null hypersurface $u=0$ (or $t=\rho$),
is an expanding sphere given by $T^2-X^2-Y^2-Z^2=0$. The solution therefore
describes a gravitational wave with an exact spherical wavefront propagating
into a Minkowski background.

It may also be observed that the metric (5--6) may be written in the form 
 $$ \d s^2= \d t^2 -\d\rho^2 -{\textstyle{1\over2}}t^2(\d z-\d\phi)^2
-{\textstyle{1\over2}}\rho^2(\d z+\d\phi)^2 $$ 
 which is clearly simply a rotation of the flat metric (7). Thus, the
space-time inside the spherical wavefront is just another version of
Minkowski space. The single soliton solution obtained from a plane symmetric
Kasner seed is a flat space-time. The only possible nonzero components of the
Weyl tensor arise from the discontinuities in the derivatives of the metric
across the wavefront. The solution described above therefore represents an
impulsive spherical wave propagating in a Minkowski background. This is in
fact a special case of the impulsive spherical gravitational wave that was
constructed by Penrose \cite{Pen72} using a ``cut and paste'' method.

It may also be observed in this case that $t=0$ simply corresponds to a
coordinate singularity in the Minkowski background. It is thus possible to add
the Minkowski region in which $0\le T<|Z|$, and then to add the time-reverse
of the solution for $T<0$ and $t<0$. The global solution then describes a
contracting impulsive gravitational wave in a Minkowski background which
collapses to a point. After the collapse, the wave then re-expands as an exact
spherical impulsive wave. The only singularity occurs at the event at which
the spherical wave has zero radius.

It may also be pointed out that Gleiser, Garate and Nicassio \cite{GlGaNi96}
have obtained a similar singularity-free solution in a one-soliton solution
generated from a Bianchi~VI$_0$ seed. In that work they describe the soliton
perturbation as ``erasing'' the ``cosmological'' singularity that occurs in
the seed.

For the case of a more general Kasner seed, solutions can be constructed as
outlined above. For $p\ne0$, they will have curvature singularities both when
$t=0$ and when $\rho=0$, so that the complete space-time is the region $t>0$,
$\rho>0$. However, these cases do not strictly have an axis at $\rho=0$ and
the space is not asymptotically flat.

\subsection {A non-soliton extension}
In the plane wave extension of \S5.1, the gravitational waves propagating in
the direction $v=$~const. simply continue into the plane wave region in which
no waves propagate in the opposite direction. However, it is natural to
consider the possible existence of gravitational waves propagating in the
extended region parallel to the shock front. Such components occur in the
soliton extension of \S5.2 in which both regions are algebraically general,
although only specific wave profiles are permitted. We now consider an
alternative extension which includes arbitrary gravitational wave components
in the extended region --- at least for the diagonal case in which such waves
have constant aligned polarization. Such an extension can be constructed as
follows.

We start by choosing coordinates in the one-soliton region with \
$f=-{1\over2}\,u^2$ \ and \ $g={1\over2}\,v^2$
\ so that $\alpha$ and $\beta$ are given by (4). We then adopt the same
expression for $\alpha(u,v)$ in the extended region so that there are no Ricci
tensor components across the shock front. The metric in the extended region
can then be taken in the form 
 $$ \d s^2=2e^{-M}\d u\,\d v-\alpha(e^V\d x^2+e^{-V}\d y^2), $$ 
 where 
 $$ V(u,v)=p\log\alpha+\tilde V(u,v) $$ 
 and $\tilde V$ is an arbitrary function satisfying $\tilde V(0,v)=0$, so that
the metric remains continuous across the shock front.

An expression for $\tilde V$ satisfying the above property is given by the
Rosen pulse solution \cite{Rosen54} which, in this context, takes the form 
 $$ \tilde V 
=\int_0^f {F(\sigma)\>\d\sigma \over\sqrt{\sigma-f}\sqrt{\sigma+g}}
=-\int_{-u^2}^0 {F(\sigma)\>\d\sigma 
\over\sqrt{\sigma+u^2}\sqrt{\sigma+v^2}}, $$ 
 where $F(\sigma)$ is an arbitrary function. In the present situation we
require $F(0)=0$ and the continuity of $F$ at $\sigma=0$ must be determined
very carefully. However, since a complete solution cannot be determined for
the solution in this form, its application here is very limited.

On the other hand, an explicit representation for $\tilde V$ for which a
complete solution can be obtained, and which satisfies the required properties
above, has been given in \cite{AleGri95}--\cite{AleGri97}. These solutions are
expressed as a sum over explicit components, each of which have the
self-similar form 
 $$ \tilde V_k(f,g)=(f+g)^k H_k\!\left({\textstyle{g-f\over f+g}}\right)  $$ 
 where $k$ is an arbitrary real parameter. Putting $\zeta=-\beta/\alpha$, the
functions $H_k(\zeta)$ satisfy the linear equation 
 $$ (\zeta^2-1) H_k''+(1-2k)\zeta H_k'+k^2 H_k = 0, $$ 
 together with the initial condition $H_k(1)=0$, so that they satisfy the
recursion relations 
 $$ H_k(\zeta) =\int_1^\zeta H_{k-1}(\zeta')\,d\zeta' \qquad
{\rm or} \qquad H_k'(\zeta)=H_{k-1}(\zeta). $$ 
 These solutions can be expressed in terms of standard hypergeometric
functions in the form 
 $$ (f+g)^k H_k\left({\textstyle {g-f\over f+g} }\right) 
= c_k\,{f^{1/2+k}\over\sqrt{f+g}}\,
F\left({\textstyle{1\over2}},{\textstyle{1\over2}}\,;
{\textstyle{3\over2}}+k\,;{f\over{f+g}}\right) $$
 where, for integer $k$, 
$c_k=(-1)^k2^k\Gamma({3\over2})/\Gamma(k+{3\over2})$. The papers 
\cite{AleGri95}--\cite{AleGri97}, however, are concerned with gravitational
waves with distinct wavefronts propagating into certain simple backgrounds.
For such situations, it is only necessary to consider cases in which
$k\ge{1\over2}$. However, in the present situation, it is necessary to choose
the minimum value of $k$ in order to remove the singularity that would
otherwise arise in the metric coefficient~$e^{-M}$, and it can be seen that
this must involve the component $k=0$.

In determining an explicit solution for the extension, we may use the fact
that 
 $$ H_0(\zeta)=\cosh^{-1}\zeta \qquad {\rm and} \qquad
H_{-1}(\zeta)={1\over\sqrt{\zeta^2-1}}. $$ 
 It is convenient initially to consider the solution $\tilde V=a_0H_0(\zeta)$,
where $a_0$ is a constant, so that 
 \begin{eqnarray}
 V &=& p\log\alpha +a_0\,H_0(\zeta) \nonumber\\
&=& p\log(f+g) +a_0\,H_0\!\left({\textstyle{g-f\over f+g}}\right). \nonumber
 \end{eqnarray}
 The remaining vacuum field equations as given in \cite{AleGri95} can then be
integrated, yielding 
 \begin{eqnarray}
 e^{-M} &=& {C|f'g'|\big(\zeta+\sqrt{\zeta^2-1}\big)^{a_0^{}p}\>
\alpha^{(p^2-a_0^2-1)/2}
\over(\zeta^2-1)^{a_0^2/2}} \nonumber\\
 &=& {C|f'g'| \left(\sqrt{g}+\sqrt{-f}\right)^{2a_0^{}p}
(f+g)^{[(a_0^{}-p)^2-1]/2}
\over(-4fg)^{a_0^2/2}}, \nonumber
 \end{eqnarray} 
 where $C$ is an arbitrary constant. With the above expressions for $f$ and
$g$, it can immediately be seen that a coordinate singularity is avoided only
if $a_0=\pm1$. Taking $a_0=1$ and a particular value for $C$, we obtain that 
 $$ e^{-M} =\left({v+u\over v-u}\right)^p (v^2-u^2)^{p^2/2}, $$ 
 indicating that the metric is now continuous as required across the shock
front. However, it may be noticed that this case in which $\tilde
V=H_0(\zeta)$ is just the diagonal case of the one-soliton solution, and
therefore belongs to the class of extensions discussed in the previous
subsection.

A more general extension in which the metric is diagonal in the extended
region can now be constructed using 
 $$  V =p\log\alpha +\sum_{n=0}^\infty a_n\, \alpha^n \,H_n(\zeta), $$ 
 where $a_0=1$ to avoid the coordinate singularity on the shock front, and the
remaining coefficients $a_n$ are arbitrary. In this case, it can be shown that 
 $$ M=-{p^2\over2}\log\alpha -p\sum_{n=0}^\infty a_n\, \alpha^n \,H_n(\zeta)
-\sum_{n=1}^\infty {1\over2n}\,\alpha^n \,K_n(\zeta) $$ 
 where 
 $$ K_n(\zeta) =\sum_{k=0}^{n-1} a_ka_{n-k} \left[
k(n-k)H_kH_{n-k}-(\zeta^2-1)H_{k-1}H_{n-k-1} \right]. $$ 
 This solution includes additional arbitrary gravitational wave components
beyond the shock front. However, we note that the curvature singularity which
occurs when $\alpha=0$ is now located on the spacelike hypersurface $u+v=0$ in
the soliton region and on the timelike hypersurface $v-u=0$ in the extended
region. Since this is a curvature singularity, it must form a boundary to the
extended space-time as illustrated in figure~6.

\section {Discussion}

In the above sections, we have considered the physical interpretation and
possible extensions for the soliton solutions with real poles in the case when
the seed solution is taken to be the Bianchi~I vacuum Kasner solution. Very
different solutions can be constructed using alternative seed solutions.
However, the character of the shock front and the possible extensions across
it is likely to have some similar properties in all cases.

At least for solitons with a Kasner seed, we have clarified the character of
the singularity that occurs on the shock front, and we have demonstrated a
number of possible extension across it. The occurrence of thin sheets of null
matter, and possible soliton extensions, have been discussed in previous
literature for some cases. We have clarified here the structure of a plane
wave extension and given a new explicit vacuum extension in which the metric
is diagonal. There are clear problems in generalising the non-soliton
extension to the non-diagonal case as the equations are then non-linear and
superposition does not apply. Thus, the full class of permissible extensions
has still not been determined.

\section* {Acknowledgments} 
The authors are very grateful to Dr G. A. Alekseev for helpful discussions on
these topics.

\section* {Appendix}

Here we consider whether it is possible to construct an exact one-soliton
solution with the same global structure as that indicated in Figure~1, but
without the presence of thin sheets of null matter. Taking the regions $I_a$
and $I_b$ to be the same, the extension would not be Kasner. Taking
$\alpha={1\over2}(v^2-u^2)$ in the one-soliton region $I_a$, it is appropriate
to consider whether it is possible to choose a gauge such that
$\alpha={1\over2}(u^2+v^2)$ in the extended region. This is clearly $C^1$
across $u=0$, so the Ricci tensor vanishes on this null hypersurface.

Essentially, we can now prove that solutions of the type outlined above do not
exist. The proof of these statements are roughly as follows: We start with the
general metric for a space-time with two spacelike hypersurface orthogonal
Killing vectors in the form 
 $$ \d s^2=2e^{-M}\d u\,\d v -\big(f(u)+g(v)\big)
\left(\chi\,\d y^2+\chi^{-1}(\d x-\omega \d y)^2\right) $$ 
 where the coefficients depend on $u$ and $v$ (or $f$ and $g$) only. The
vacuum field equations imply that 
 \begin{equation}
 e^{-M}={|f'g'|\over\sqrt{f+g}}e^{-S}, 
 \end{equation}
 where 
 \begin{equation}
 S_f=-{\textstyle{1\over2}}(f+g){({\chi_f}^2+{\omega_f}^2)\over\chi^2},
\qquad S_g=-{\textstyle{1\over2}}(f+g){({\chi_g}^2+{\omega_g}^2)\over\chi^2}.
 \end{equation}

We may now adopt a gauge such that 
 $$ f(u)={\textstyle{1\over2}}\epsilon u^2, \qquad
g(v)={\textstyle{1\over2}}v^2. $$ 
 There is no loss of generality in adopting this form for $g(v)$ and the
freedom $u\to u'(u)$ has been used to obtain the simplest expression for which
$f(0)=0$, and $f'(0)=0$ to avoid any null matter on $u=0$. The freedom in
rescaling $u$ can be further used to set $\epsilon=\pm1$. In this case we have 
 $$ f'=\epsilon u. $$ 
 Thus (8) implies that, for $M$ to be continuous across the front $u=0$ (on
which $f+g>0$), $S$ must contain the term $\log u$. i.e. near the front
$u=0$, $S$ must behave as 
\begin{eqnarray}
 S &\sim& \log u + {\rm const.} +\dots \nonumber \\
 &\sim& {\textstyle{1\over2}}\log|f| +\dots \nonumber 
\end{eqnarray}
 Thus 
 $$ S_f \sim {1\over2f} +\dots $$ 
 It can then be seen that the first equation in (9) can only be satisfied
near the wavefront if $f<0$. i.e. it is necessary that 
 $$ \epsilon=-1. $$ 
 Thus, solutions with $\alpha=u^2+v^2$ do not exist near $u=0$. Further, for a
vacuum extension, it is always possible to choose a gauge such that
$\alpha=v^2-u^2$.

\end{document}